\documentclass{article}
\usepackage[affil-it]{authblk}
\usepackage{graphicx}
\usepackage{amsmath}
\usepackage[latin1]{inputenc}
\usepackage{listings}
\usepackage{framed}
\usepackage{xcolor}
\colorlet{shadecolor}{gray!20}

\lstnewenvironment{mat}
{\lstset{language=mathematica,mathescape,columns=flexible}}
{}
\begin{document}
 \title{Comment on ``Exact Analytic Second Virial Coefficient for the 
Lennard-Jones Fluid'' (arXiv:0909.3326v1)}
 \author{Isidro Cachadiña Gutiérrez 
 \thanks{Electronic address: \texttt{icacha@unex.es}}
 }
 
 \affil{Departamento de Física Aplicada\\
 Universidad  de  Extremadura\\
 Avda. de Elvas, s/n\\
 06006 Badajoz (Spain)}
 
\maketitle
 \begin{abstract}
   The asymptotic expansion for $T\rightarrow 0$ from Byung Chan Eu of 
   $B_2(T)=-16\sqrt{2\pi}v_0e^{\varepsilon\beta}(\varepsilon\beta)^{3/2}
   \biggl[ 1+\dfrac{19}{16\epsilon\beta}+\dfrac{105}{512(\epsilon\beta)^2}\dots 
\biggr]$ is wrong.  The correct expression is 
$B_2(T)=-2\sqrt{2\pi}v_0e^{\varepsilon\beta}(\varepsilon\beta)^{-1/2}\biggl[
1+\dfrac{15}{16\varepsilon\beta}+\dfrac{945}{512(\varepsilon\beta)^2}
+\dots\biggr]$.
\\[2cm]
 \end{abstract}

Byung Chan Eu gave an exact analytical solution for the second virial 
coefficient valid for the entire range of temperature which reads as:
\begin{equation}
 \begin{array}{ll}  
-\dfrac{B_2(x)}{v_0\sqrt{2}x^{1/4}} =  & 
 4\Gamma\biggl(\dfrac{3}{4}\biggr)\biggl[6xM\biggl(\dfrac{7}{4},
\dfrac{3}{2},x\biggr)-(1+4x)M\biggl(\dfrac{3}{4},
  \dfrac{1}{2},x\biggr)\biggr]\\
  & +\biggl[
   2\Gamma\biggl(\dfrac{1}{4}\biggr)\sqrt{x}\biggl[
   \dfrac{10}{3}xM\biggl(\dfrac{9}{4},\dfrac{5}{2},x\biggr)+
   (1-4x)M\biggl(\dfrac{5}{4}, \dfrac{3}{2},x\biggr)
  \biggr].
 \end{array}
\end{equation}
where $x=\varepsilon\beta=\varepsilon/kT$, $v_0=\pi\sigma^3/6$, and $M(a,b,x)$, 
is a confluent hypergeometric function of Krummer \cite{Abramowitz}:
\begin{equation}
  M(a,b,x) = \sum_{n=0}^\infty \dfrac{(a)_nx^n}{(b)_n n!},
\end{equation}
where $(a)_n$ is the Pochhammer symbol \cite{Abramowitz}
\begin{equation}
 \begin{array}{ll}
 (a)_0 & =1, \\
 (a)_n &= a(a+1)(a+2)\dots (a+n-1) \qquad (n\geq 1).\\
 \end{array}
\end{equation}
The asymptotic behaviour for $T\rightarrow 0$ can be 
determined using the following asymptotic expansion of $M(a,b,x)$ for real 
$x$ \cite{Abramowitz} when $x\rightarrow \infty$:
\begin{equation}
 M(a,b,x)=\dfrac{\Gamma(b)}{\Gamma(a)}e^x x^{a-b}\biggl[\sum_{n=0}^{m-1} 
\dfrac{(b-a)_n(1-a)_n}{n!}x^{-n}+O(x^{-m})\biggr].
\end{equation}

Byung Chan Eu claimed that the following expression 
\begin{equation}
 B_2^{Chan}(T)=-16\sqrt{2\pi}v_0e^{\varepsilon\beta}(\varepsilon\beta)^{3/2}
\biggl[1+\dfrac{19}{16\varepsilon\beta}+\dfrac{105}{
512(\varepsilon\beta)^2}+\cdots\biggr]
\label{Chan-approx}
\end{equation}
is the asymptotic behaviour of the second virial coefficient for 
$T\rightarrow 0$.

\begin{figure}[h]
 \begin{center}
  \includegraphics[width=0.7\linewidth]{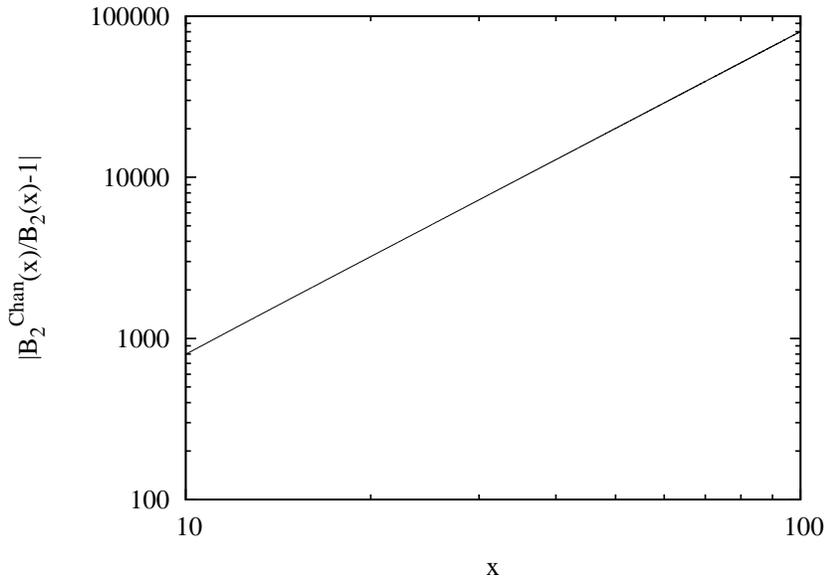}
 \end{center}
 \caption{Absolute relative deviation of the asymptotic expansion 
proposed by B. Chan Eu for the second virial coefficient. This curve have to 
tend to zero for larger values of $x$, but on the contrary, it grows as $x^2$.}
 \label{fig-01}
\end{figure}

We can see easily that this expansion is wrong by calculating 
\begin{equation}
 \biggl|\dfrac{B_2^{Chan}(x)}{B_2(x)}-1\biggr|,
 \label{Remainder-Chan}
\end{equation}
and seeing that the above expression doesn't go to 
zero for larger values of $x$.  When this function is plotted (as it is done in 
Fig. \ref{fig-01})  one can see that ratio between $|B_2^{Chan}(x)/B_2(x)-1|$ 
grows as $x^2$ and therefore $B_2^{Chan}(x)$ is not the asymptotic expansion of 
$B_2(x)$. 

With the aid of the Mathematica  program \cite{Mathematica8}  the asymptotic 
expansion can be easily calculated (see Appendix 1) to give:
\begin{equation}
B_2(x)=-2\sqrt{2\pi}v_0e^{\varepsilon\beta}(\varepsilon\beta)^{-1/2}\biggl
[
1+\dfrac 
{15}{16x}+\dfrac{945}{512x^2}+\dfrac{45045}{
8192 x^3}+\dots\biggr].
\label{my-expansion}
\end{equation}
The factor $x^2$ that can be seen in Fig. \ref{fig-01} is due the ratio of 
$x^{3/2}/x^{-1/2}$ between the expression here deduced and the one from Byung 
Chan Eu.

\begin{figure}[h]
 \begin{center}
  \includegraphics[width=0.7\linewidth]{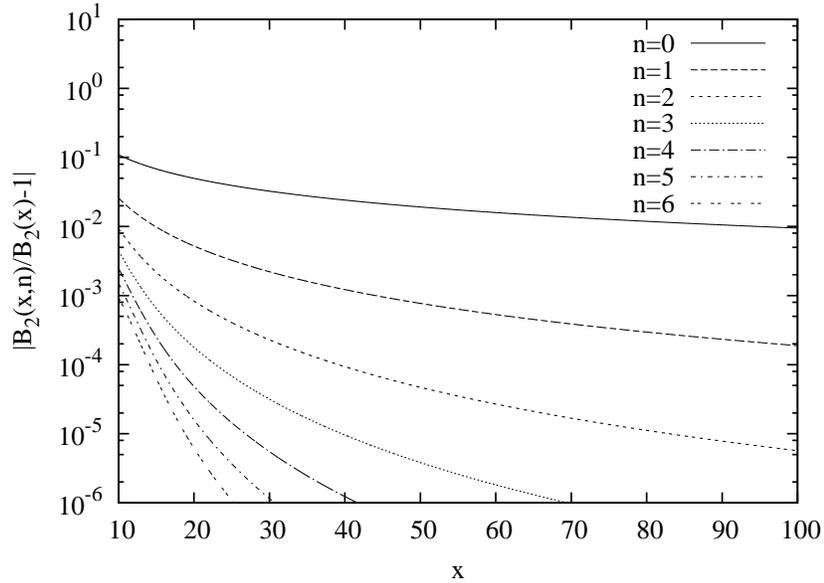}
 \end{center}
 \caption{Absolute relative deviation of the asymptotic expansion, $B_2(x,n)$ 
(Eq. (\ref{my-expansion-n})), as a function of the number of terms $n$ included. 
}
 \label{fig-02}
\end{figure}

In order to check the convergence Eq. (\ref{my-expansion}) I plotted in Fig. 
\ref{fig-02} this expansion taking different number of terms in the polynomial 
of $x^{-1}$.  Thus, calling:
\begin{equation}
 B_2(x,n)=-2\sqrt{2\pi}v_0e^{\varepsilon\beta}(\varepsilon\beta)^{-1/2}
  \sum_{i=0}^n \dfrac{a_i}{x^i}
\label{my-expansion-n}
\end{equation}
where:
\begin{equation}
 \begin{array}{ll}
  a_0 & =1 \\
  a_1 & = 15/16\\
  a_2 & = 945/512\\
  a_3 & = 45045/8192\\
  a_4 & = 11486475/524288\\
  a_5 & = 916620705/8388608\\
  a_6 & = 175685635125/268435456\\
  a_7 & = 19651693186125/4294967296\\
  a_8 & = 20103682129405875/549755813888\\
 \end{array}
\end{equation}
we can check how many terms are needed to achieve the desired accuracy.  For 
example, taking $n=6$ the expansion of Eq. (\ref{my-expansion-n} gives 6 exact 
digits when $x>25$.

\section*{Appendix}
This is the Mathematica \cite{Mathematica8} notebook I used to calculate the 
asymptotic expansions.

\begin{mat}
   
In[1]:= 
(* True value of -B2/v0 *)

B2[x_] :=  Sqrt[2]*x^(1/4)*(4*Gamma[3/4]*(6*x*Hypergeometric1F1[7/4, 3/2, x] - 
  (1 + 4*x)*Hypergeometric1F1[3/4, 1/2, x]) + 2*Gamma[1/4]*     
  Sqrt[x]*(10/3*x*Hypergeometric1F1[9/4, 5/2, x] + (1 - 4*x)*   
  Hypergeometric1F1[5/4, 3/2, x]))

(* Check if TB^-1 = 0.2925 *)

In[2]:= FindRoot[B2[x] == 0, {x, 0.3}]

Out[2]= {x -> 0.292575}

In[3]:= (* Asymptotic expansion of Krummer functions up to n *)

M[a_, b_, x_, n_] := 
  Gamma[b]/Gamma[a]*Exp[x]*x^(a - b)*
  Sum[Pochhammer[b - a, k]*Pochhammer[1 - a, k]*x^-k/k!, {k, 0, n}]


In[5]:= (* The value of -B2/v0 using the asymptotic expansion up to term n *)

F[x_, n_] :=  Sqrt[2]*x^(1/4)*(4*
     Gamma[3/4]*(6*x*M[7/4, 3/2, x, n] - (1 + 4*x)*M[3/4, 1/2, x, n]) +
    2*Gamma[1/4]*Sqrt[x]*(10/3*x*M[9/4, 5/2, x, n] 
    + (1 - 4*x)*M[5/4, 3/2, x, n]))


In[6]:= (* The two first approximations of F[x,n] are zero *)

FullSimplify[F[x, 0]]

Out[6]= 0

In[7]:= FullSimplify[F[x, 1]]

Out[7]= 0

In[8]:= Ap2[x_] = FullSimplify[F[x, 2]]

Out[8]= (2 E^x Sqrt[2 \[Pi]])/Sqrt[x]


In[9]:= Apn[x_, n_] := 
 Expand[FullSimplify[F[x, n]/F[x, 2]]]*FullSimplify[F[x, 2]]

In[10]:= Apn[x, 7]

Out[10]= (2 E^x Sqrt[
 2 \[Pi]] (1 + 916620705/(8388608 x^5) + 11486475/(524288 x^4) + 
   45045/(8192 x^3) + 945/(512 x^2) + 15/(16 x)))/Sqrt[x]

In[11]:= CoefficientList[Expand[FullSimplify[F[x, 10]/F[x, 2]]], 1/x]


Out[11]= {1, 15/16, 945/512, 45045/8192, 11486475/524288, \
916620705/8388608, 175685635125/268435456, 19651693186125/4294967296, \
20103682129405875/549755813888}

\end{mat}

\end{document}